\newcounter{myctr}
\def\myitem{\refstepcounter{myctr}\bibfont\noindent\ifnum\themyctr>9\else\phantom{0}\fi\hangindent17pt\themyctr.\enskip}

\documentclass{ws-ijqi}

\usepackage{makeidx}
\usepackage{amsfonts}

\newcommand{\hr}{{\mathcal{H}}}
\newcommand{\qed}{{\hfill $\square$}}
\newcommand{\idn}{\mathbf{1}}
\newcommand{\C}{{\mathbb C}}

\begin{document}

\markboth{Markus M\"uller}
{On the Quantum Kolmogorov Complexity of Classical Strings}

\catchline{}{}{}{}{}

\title{ON THE QUANTUM KOLMOGOROV COMPLEXITY OF CLASSICAL STRINGS}

\author{MARKUS M\"ULLER}

\address{Max Planck Institute for Mathematics in the Sciences, Inselstr. 22,\\
04103 Leipzig, Germany, {\em and}\\
Institute of Mathematics, TU Berlin, Stra\ss e des 17. Juni 136,\\
10623 Berlin, Germany\\
mueller@math.tu-berlin.de}

\maketitle

\begin{history}
\received{Day Month Year}
\revised{Day Month Year}
\end{history}

\begin{abstract}
We show that classical and quantum Kolmogorov complexity of binary strings agree up to an additive constant.
Both complexities are defined as the minimal length of any (classical resp. quantum) computer program that outputs
the corresponding string.

It follows that quantum complexity is an extension of classical complexity to the domain of quantum states.
This is true even if we allow a small probabilistic error in the quantum computer's output.
We outline a mathematical proof of this statement, based on an inequality for outputs of quantum operations and a classical program
for the simulation of a universal quantum computer.
\end{abstract}


\section{Introduction}
Are quantum computers more powerful than classical ones? Concerning computational complexity, i.e.
the speed of solving certain problems, the answer seems to be yes. A well-known example
is Shor's algorithm (cf. Ref.~\refcite{NielsenChuang}) for factoring large integers on quantum computers in polynomial time,
which is generally believed to be impossible for a classical computer.

In this paper, we intend to compare classical and quantum computers with respect to a different complexity
measure, which does not care about the time of computation, but instead measures the minimal description length.
More in detail, let $\{0,1\}^*=\{\varepsilon,0,1,00,01,\ldots\}$ denote the finite binary strings, where $\varepsilon$ is
the empty string, and let $U$ be a universal computer. Then, the Kolmogorov complexity of any string $x\in\{0,1\}^*$ is defined as
\begin{equation}
   C(x):=\min\{\ell(p)\,\,|\,\,U(p)=x\},
   \label{eqDefC}
\end{equation}
i.e. the length of the shortest computer program that makes $U$ compute and output $x$. Since it was developed
in the 1960's by Solomonoff, Kolmogorov and Chaitin, this complexity measure has turned out to have useful applications in several
different areas of mathematics (as explained e.g. in Ref.~\refcite{Vitanyibook}), besides being an interesting object of study itself.

Due to the rapid development of quantum information theory,
there has been some interest in recent years (cf. Refs.~\refcite{Berthiaume}, \refcite{Vitanyi}, \refcite{Gacs})
to extend this quantity to the quantum case.
While inputs and outputs of classical computers are binary strings, quantum computers may also work with
superpositions of strings. Thus, it makes sense to consider quantum states like, for example,
\[
   |\psi\rangle=\frac 1 {\sqrt 2}\left(\strut |01\rangle+|10\rangle\right)
\]
as inputs and outputs of quantum computers, and to assign a quantum Kolmogorov complexity measure $QC(|\psi\rangle)$ to those
states, by a definition similar to Eq.~(\ref{eqDefC}). Such a definition has first been given by Berthiaume,
van Dam, and Laplante in 2001 in Ref.~\refcite{Berthiaume}.

Formally, the aforementioned quantum state $|\psi\rangle$ is an element of some Hilbert space $\hr_{\{0,1\}^*}$ that contains the classical
binary strings as an orthonormal basis. We do not restrict our considerations to pure states like $|\psi\rangle$, but
also allow mixed states\footnote{It is natural to allow mixed states also, because prefixes of (pure) qubit strings
can be mixed, as explained in Ref.~\refcite{MuellerRogers}.}, i.e.
arbitrary density operators $\rho$ on $\hr_{\{0,1\}^*}$ as inputs and outputs of quantum computers.
We call these density operators {\em qubit strings}.

Once a quantum analogue of Kolmogorov complexity has been defined, the question
arises how classical and quantum complexity are related to each other on their common domain of definition.
In other words, how are $C(x)$ and $QC(|x\rangle)$ related for classical binary strings $x\in\{0,1\}^*$?
Generically, it might be true that quantum computers offer better possibilities to find short (quantum)
descriptions of binary strings than classical computers. In this case, we would have $QC(|x\rangle)\ll C(x)$ for some strings $x$.

Here, we shall show that this is not the case, and that $C(x)$ and $QC(|x\rangle)$ are equal up to an additive constant.
We prove an analogue of this result also for the case that a certain probabilistic error is allowed for the quantum
computer's output (corresponding
to the complexity notion $QC^\delta$ introduced below in Section~\ref{SecQComplexity}).
Our result answers an open problem posed in Ref.~\refcite{Berthiaume}.

Note that if $|\psi\rangle$ is a non-classical qubit string, then the classical complexity $C(|\psi\rangle)$ is undefined,
because $|\psi\rangle$ is not a classical binary string. However, if we define $C(|\psi\rangle)$ to be the {\em shortest classical description}
of $|\psi\rangle$, then $C$ and $QC$ behave completely differently: For example, if
\[
   |\psi\rangle:=\sqrt{\Omega}|0\rangle+\sqrt{1-\Omega}|1\rangle,
\]
where $0<\Omega<1$ denotes the (non-computable) halting probability of a universal Turing machine (cf. Ref.~\refcite{Vitanyibook}),
then $|\psi\rangle$ is a quantum description of itself (with a length of one qubit), whereas $|\psi\rangle$
does not even possess a finite classical description (i.e. computer program to compute $|\psi\rangle$).
Even an approximate description must contain a large number of bits, as $\Omega$ is a random real number.
Hence $QC(|\psi\rangle)\ll C(|\psi\rangle)$.
Thus, the result in this paper does {\em not} imply that $C$ and $QC$ are just the same; they only agree
on the domain of classical strings.

We remark that the exposition in this paper focuses more on a compact presentation than on mathematical rigidity.
For more mathematical details of the proof, we refer the reader instead to Ref.~\refcite{MeineDiss}.

\section{Quantum Kolmogorov Complexity}
\label{SecQComplexity}
We would like to define the quantum complexity of some qubit string $\rho$ as the length of
the shortest quantum computer program that produces $\rho$. But while classical strings are discrete objects,
the set of qubit strings is a continuum. If two qubit strings are close to each other, they are very difficult
to distinguish by means of any quantum measurement. Thus, it does not make sense
to demand that the desired output $\rho$ is produced perfectly by the computer program, but it is more useful to allow a certain error tolerance.

To quantify the difference between two qubit strings, it is natural to use the trace distance (see Ref.~\refcite{NielsenChuang}),
which is defined as $\|\rho-\sigma\|_{\rm Tr}:=\frac 1 2 {\rm Tr}|\rho-\sigma|$.

Quantum Kolmogorov complexity can now be defined in essentially two ways: First, one can just fix some error tolerance
$\delta>0$.	Second, one can demand that the quantum computer outputs the qubit string $\rho$ as accurately as one wants,
by supplying the machine with a second parameter as input that represents the desired accuracy.
We consider both approaches and
follow the lines of Berthiaume et al. (Ref.~\refcite{Berthiaume}) except for slight modifications:

\begin{definition}[Quantum Kolmogorov Complexity]
\label{DefQC}
Let $U$ be a universal quantum computer and $\delta>0$. Then, for every qubit string $\rho$, we define
\[
   QC^\delta(\rho):=\min\{\ell(\sigma)\,\,|\,\,\|\rho-U(\sigma)\|_{\rm Tr}\leq\delta\}.
\]
Moreover, we set
\[
   QC(\rho):=\min\left\{\ell(\sigma)\,\,\left|\,\,\|\rho-U(\sigma,k)\|_{\rm Tr}\leq\frac 1 k\right.\mbox{ for every }k\in\mathbb{N}\right\}.
\]
\end{definition}
The universal quantum Turing machine (QTM), constructed in Ref.~\refcite{BernsteinVazirani} and modified for the definition of Kolmogorov complexity
with respect to base length in Ref.~\refcite{StronglyUniversal}, can serve as a model for a universal quantum computer. 

The length $\ell(\sigma)$ of a qubit string $\sigma$ is here defined as the maximal length of any classical
string that has non-zero overlap with $\sigma$. For example, the qubit string $|\psi\rangle:=
\frac 4 5 |00\rangle-\frac 3 5 |1011\rangle$ (or rather its
corresponding density operator $|\psi\rangle\langle\psi|$) is understood to have length $\ell(|\psi\rangle)=4$.
This is called the ``base length'' by Bostr\"om and Felbinger in Ref.~\refcite{BostroemFelbinger}.

Quantum Kolmogorov complexity shares many properties with its classical counterpart. For example,
there is a constant $c\in\mathbb{N}$ such that
for every qubit string $\rho$, it holds $QC(\rho)\leq \ell(\rho)+c$ (the same for $QC^\delta$), and the
value of $QC(\rho)$ depends on the choice of the universal quantum computer $U$ only up to an additive constant
as shown in Ref.~\refcite{StronglyUniversal}.

\section{Quantum Kolmogorov Complexity of Classical Strings}
The main result of this paper is the following theorem:
\begin{theorem}[Main Theorem]\label{MainTheorem}
For every classical string $x\in\{0,1\}^*$,
\begin{equation}
   C(x)=QC(|x\rangle)+\mathcal{O}(1),
   \label{eqO1}
\end{equation}
i.e. the absolute value of the difference of $C$ and $QC$ is bounded by a constant on the
domain of classical strings. Moreover, for every rational $0<\delta<\frac 1 {2e}$, there
are constants $c_\delta,c_{\delta}'\in\mathbb{N}$ such that
\[
   QC^\delta(|x\rangle)\leq C(x)+c_{\delta}\leq \frac {QC^\delta(|x\rangle)} {1-4\delta} +c_\delta'.
\]
\end{theorem}
Thus, the only possible difference between classical and quantum complexity for classical strings is the factor
$\frac 1 {1-4\delta}>1$ in the second equation, indicating that a quantum computer with large
error tolerance $\delta$ might sometimes have shorter programs for generating classical strings
than a classical computer (which always has error tolerance zero).

Note that in a very interesting paper based on a different complexity notion (Ref.~\refcite{Gacs}), G\'acs has already shown
a prefix-free analogue of Eq. (\ref{eqO1}).

\subsection{Outline of the Proof of Theorem~\ref{MainTheorem}}
We start with a simple argument, showing that for every rational number $\delta>0$, there is a constant $k_\delta\in\mathbb{N}$ such that
\[
   QC^\delta(\rho)\leq QC(\rho)+k_\delta
\]
for every qubit string $\rho$: This is clear, since every (quantum) program that computes $\rho$ to any desired accuracy
can be transformed into a program that computes $\rho$ within some fixed accuracy $\delta$. We just have to additionally
specify the parameter $\delta$, which costs at most a fixed number of bits $k_\delta$.

Also, we claim that there is some constant $c\in\mathbb{N}$ such that for every classical string $x$, it holds
\[
   QC(|x\rangle)\leq C(x)+c.
\]
This can be seen as follows: As Bennett has shown in Ref.~\refcite{Bennett}, we can choose the classical computer which is used in the
definition of $C(x)$ to be reversible.
But every reversibe computer is also a (special case of a) quantum computer and can thus be simulated by our universal
quantum computer $U$.
We can thus find a program for $U$ which computes $x$ perfectly, and we only have to add a constant number
of bits for the simulation.

It remains to show that there is some constant $k_\delta''\in\mathbb{N}$ such that
\begin{equation}
   C(x)\leq\frac 1 {1-4\delta} QC^\delta(|x\rangle)+k_\delta''.
   \label{eqDesired}
\end{equation}
We prove this inequality by giving a classical computer program $p$ that
simulates the universal quantum computer $U$.\footnote{Note that there is no problem in simulating a quantum
computer (e.g. a QTM) on a classical computer,
if a classical description of that QTM and of the input qubit string is given.
The simulation will be very inefficient, i.e. very slow, but the time of computation is irrelevant for Kolmogorov complexity.}
More in detail, on input $i,n\in\mathbb{N}$ and $0<\delta\in\mathbb{Q}$,
that program approximately computes the $i$-th classical string that is generated within some accuracy $\delta$ by the
quantum computer $U$
on any input qubit string of length $n$. It works as follows:
\begin{itemize}
\item[1.] Set the time $t:=0$ and the counter $c:=0$.
\item[2.] Compute a discretization of the set of qubit strings (density matrices) of length less than or equal to $n$, that is,
the description of a finite set $\{\sigma_1,\ldots,\sigma_M\}$ of qubit strings such that for every qubit string $\sigma$ of
length at most $n$, there is some $j\in\{1,\ldots,M\}$ such that $\sigma$ is very close to $\sigma_j$ in trace distance.
\item[3.] For every $j\in\{1,\ldots,M\}$, simulate the quantum computer $U$ on input $\sigma_j$ for $t$ time steps and decide
whether $U$ approximately halts on input $\sigma_j$ at time $t$. If this is the case, then compute a good
approximation of the corresponding output, and check for every classical string $x$ of length $\ell(x)\leq t$
whether that output is very close to $|x\rangle$ (with a suitable distance threshold depending on $\delta$).
If such a string $x$ is found, increase the counter by one,
i.e. $c:=c+1$, and check if $c=i$. If this is true, then output $x$ and halt.
\item[4.] Increase the time $t$ by one, i.e. $t:=t+1$, and go back to step 3.
\end{itemize}

If every approximation in every step is done in appropriate accuracy, then this program will find every classical string $x$
which is generated by some input qubit string $\sigma$ of length $\ell(\sigma)=n$ within accuracy $\delta$,
whenever the corresponding index $i$ is given as input. But then, we can construct a classical computer
program $p_x$ to produce $x$: We just have to append a description of the program above to a description
of the integer $i$. The resulting program $p_x$ will have length
\[
   \ell(p_x)=k'_\delta+\lceil\log_2 N_{n,\delta}\rceil,
\]
where $N_{n,\delta}$ denotes an upper bound on the possible values of $i$ (given $n$ and $\delta$), i.e. $i\leq N_{n,\delta}$,
and $k'_\delta$ is a constant that depends on $\delta$ (since
a description of $\delta$ has to be included in the program).

As long as the string $x$ satisfies $QC^\delta(|x\rangle)\leq n$, the program $p_x$ must terminate and return output $x$, which can
be seen as follows. If $QC^\delta(|x\rangle)\leq n$, then there exists an input qubit string $\sigma$ such that $U(\sigma)$
is $\delta$-close in trace distance to $|x\rangle\langle x|$, and there is a corresponding halting time $\tau$. If the counter $t$
in the four-step program given above is large enough, it will reach $\tau$ and finally discover the corresponding output string $x$.

To get a useful number of bits $\lceil \log_2 N_{n,\delta}\rceil $ to encode the value of $i$,
we have to estimate the total number $\tilde N_{n,\delta}$ of classical strings
that are produced by the quantum computer $U$ on some input of length less than or equal
to $n$, if some error tolerance $\delta$ is allowed.
(Since it may be hard to compute the actual value of $\tilde N_{n,\delta}$, we do not use $N_{n,\delta}=\tilde N_{n,\delta}$ directly;
we only have to ensure that $N_{n,\delta}\geq \tilde N_{n,\delta}$).
Viewed as qubit strings, the classical strings are mutually orthogonal. Moreover, it has been shown in Ref.~\refcite{QBrudno}
that the quantum computer $U$ as a mapping
on the qubit strings is a quantum operation, as every physically realizable operation on quantum
states (cf. Ref.~\refcite{NielsenChuang}). Thus, we have an instance of a more general problem:
Given some quantum operation on some Hilbert space, how many mutually orthogonal vectors can be created by
that operation if some error $\delta$ is allowed? We give an estimate below in Lemma~\ref{LemQuantumCountingArgument}.
Using the result of that lemma, we get a constant $k'''_\delta\in\mathbb{N}$ such that
\[
   \log_2 \tilde N_{n,\delta}\leq 
      \frac{n+1+4\delta\log_2\frac 1 \delta}
      {1-4\delta}\leq \left\lceil \frac n {1-4\delta}\right\rceil + k'''_\delta,
\]
since the qubit strings of length $n$ or less are density matrices on the Hilbert space $\hr$ which is
spanned by the classical strings of length less than or equal to $n$, and thus $d:=\dim\hr=2^{n+1}-1$.
Thus, it is possible to choose the fixed value $\log_2 N_{n,\delta}:=\left\lceil \frac n {1-4\delta}\right\rceil + k'''_\delta$ as
the number of bits that are used in the description of $p_x$ to specify the corresponding index $i$.
The program $p_x$ will then have length
\[
   \ell(p_x)=k'_\delta+\left\lceil \frac n {1-4\delta}\right\rceil + k'''_\delta.
\]
In particular, the value of $n$ can be deduced from the length of $p_x$. This is the reason why we do {\em not} have
to implement an additional description of $n$ in the program $p_x$ (which would add an additional number of about
$\log n$ bits). Instead, the program $p_x$ may compute $n$ from its own length, more in detail, from the length
of the description of $i$.\footnote{The part of $p_x$ which does {\em not} contain the description of $i$ might be
encoded, for example, in some self-delimiting manner, e.g. by means of a prefix-free code, such that it becomes clear at
what bit position the description of $i$ starts. The description of $i$, however, does {\em not} need to be self-delimiting,
since programs may always determine their own lengths: in the Turing-machine picture, they may look for the first blank
symbol on the tape, for example.}

We get a bound on the classical Kolmogorov complexity of $x$ via
\begin{equation}
   C(x)\leq \ell(p_x)
   \leq k''_\delta+\frac n {1-4\delta},
   \label{EqForcing}
\end{equation}
where $k''_\delta:=k'_\delta+k'''_\delta+1$. The desired inequality (\ref{eqDesired}) now follows from the
special case $n:=QC^\delta(|x\rangle)$ (we have only used ``$\geq$'' in the calculation above).

Finally, it remains to prove that
\begin{equation}
   C(x)\leq QC(|x\rangle)+c.
   \label{eqDesired2}
\end{equation}
We start by bounding the number $N$ of classical strings $s$ with quantum Kolmogorov complexity $QC(|s\rangle)\leq n$. First notice
that if $k\in\mathbb{N}$ is fixed, then the map $\sigma\mapsto U(\sigma,k)$ is a quantum operation (cf. Ref.~\refcite{MeineDiss}).
Thus, if we define
\[
   S_k^{(n)}:=\left\{ s\in\{0,1\}^* \,\,|\,\, \exists \sigma:\ell(\sigma)\leq n, \| U(\sigma,k)-|s\rangle\langle s\|_{\rm Tr}
   \leq \frac 1 k\right\},
\]
then it follows from Lemma~\ref{LemQuantumCountingArgument} that the cardinality of these sets is bounded by
\begin{equation}
   \log_2 \# S_k^{(n)} \leq \frac{n+1+\frac 4 k \log_2 k}{1-\frac 4 k}.
   \label{EqBounded3}
\end{equation}
Consequently, if a classical string $x$ has quantum Kolmogorov complexity $QC(|x\rangle)\leq n$, then it must be
an element of $S_k^{(n)}$ for every $k$, i.e. $x\in\bigcap_{k\in\mathbb{N}} S_k^{(n)}$ (cf. the definition of $QC$ in
Definition~\ref{DefQC}). Hence the number $N$ of such strings is bounded by
\begin{equation}
   \log_2 N\leq \log \inf_{k\in\mathbb{N}} \# S_k^{(n)} \leq \inf_{k\in\mathbb{N}} \frac{n+1+\frac 4 k \log_2 k}{1-\frac 4 k}=n+1.
   \label{EqLog2N}
\end{equation}
It also follows from inequality~(\ref{EqBounded3}) that there is some $K\in\mathbb{N}$ (possibly depending on $n$) such that
$\# S_k^{(n)}<2^{n+1}+1$ for every $k\geq K$. Since cardinalities are integer-valued, this means
that $\# S_k^{(n)}\leq 2^{n+1}$ for every $k\geq K$.

In the proof of inequality~(\ref{eqDesired}), we have given a classical computer program $p_x$ that approximately\footnote{Note
that we do not take into account here all the possible numerical errors that the program has to deal with.
See Ref.~\refcite{MeineDiss} for a more detailed (hence much more complicated) discussion concerning how to choose the 
accuracies in the approximations and numerical computations.
} computes the $i$-th classical binary string $x$ that is generated by some quantum computer program of length less than $n$
up to trace distance $\delta$. We modify this program to obtain a new program $\tilde p_x$,
containing an integer $i\in\mathbb{N}$, which works as follows:
\begin{itemize}
\item[1.] Set $\delta:=\frac 1 6$.
\item[2.] Approximately compute the list of all classical strings $s$ that are generated by any quantum computer program of length less
than or equal to $n$ up to trace distance $\delta$.
\item[3.] If this list contains more than $2^{n+1}$ items, then let $\delta:=\frac \delta 2$ and go back to step 2.
\item[4.] Output the $i$-th element of this list.
\end{itemize}
This program always terminates: if $\delta$ is small enough, then $k:=\lceil \frac 1 \delta\rceil\geq K$ holds, and
the set $S_k^{(n)}$ contains at most $2^{n+1}$ elements, which interrupts the loop in step 3.
Due to the calculations above, every classical string $x$ with $QC(|x\rangle)\leq n$ is output by this computer
program $\tilde p$ if $i$ is chosen appropriately. Moreover, it holds $i\leq N\leq 2^{n+1}$ due to inequality~(\ref{EqLog2N}), such that
it is possible to encode $i$ in $n+1$ bits.
In contrast to the program $p_x$, this program $\tilde p_x$ does not need any specific parameter $\delta$ as input.
Moreover, it will also compute $n$ directly from its own length similarly as explained above for the program $p_x$. It follows that
there is a constant $c'\in\mathbb{N}$ (and $c:=c'+1$) such that
\[
   C(x)\leq \ell(\tilde p_x)= c'+n+1=n+c.
\]
This proves inequality~(\ref{eqDesired2}).
Combining all the previous inequalities completes the proof of the theorem.
\qed

\section{An Inequality for the Almost-Output Dimension of Quantum Operations}
Here is an analytical result that we have used above to estimate the number of classical outputs.
Note that a special case of this lemma has been published in Ref.~\refcite{QBrudno}. For $\delta=0$,
setting $0\log\frac 2 0 :=0$, that lemma states that quantum operations cannot increase
the number of dimensions, which is obvious, since quantum operations are linear maps.
For small $\delta>0$, the number of orthonormal vectors can increase only slightly.

To state our result, we use the trace norm $\|\rho-\sigma\|_{\rm Tr}:=\frac 1 2 \rm{Tr}|\rho-\sigma|$,
which is used also in Ref.~\refcite{NielsenChuang}. It is easy to translate the result to the $1$-norm, since
$\|\cdot\|_{\rm Tr}=\frac 1 2 \|\cdot\|_1$. In the field of classical Kolmorogov complexity,
a ``counting argument'' is frequently used, stating that $N$ different inputs for
a computer can have at most $N$ different outputs. Since our lemma is some kind of quantum generalization
of this result, we call it a ``quantum counting argument''.

The proof is based on Holevo's $\chi$-quantity associated to any ensemble $\mathbb{E}_\rho
:=\left\{\lambda_i,\rho_i\right\}_i$, consisting of probabilities $0\leq\lambda_i\leq 1$,
$\sum_i \lambda_i=1$, and of density matrices $\rho_i$ acting on a Hilbert space $\hr$.
Setting $\rho:=\sum_i\lambda_i\rho_i$, the $\chi$-quantity is defined as follows:
\[
   \chi(\mathbb{E}_\rho):=S(\rho)-\sum_i \lambda_i S(\rho_i)=\sum_i\lambda_i S(\rho_i,\rho),
\]
where $S(\cdot,\cdot)$ denotes the relative entropy.
\begin{lemma}[Quantum Counting Argument]
\label{LemQuantumCountingArgument}
Let $\hr$ and $\hr'$ be separable Hilbert spaces with $0<d:=\dim\hr<\infty$, and let $0\leq\delta<\frac 1 {2e}$.
If $\mathcal{E}$ is a quantum operation from the density operators of $\hr$ to those of $\hr'$,
then the maximal number $N$ of mutually orthonormal vectors on $\hr'$ which are all produced by $\mathcal{E}$
within trace distance $\delta$ on some (possibly mixed) input is bounded by
\[
   \log_2 N\leq \frac{\log_2 d + 4\delta\log_2\frac 1 \delta}{1-4\delta}.
\]
\end{lemma}
{\bf Proof.}
For $\delta=0$, the assertion of the theorem is trivial (setting, as usual, $0\log\frac 1 0:=0$), so
assume $\delta>0$. Let $N_\delta$ be a set of orthonormal vectors $|\psi\rangle\in\hr'$ such that
there exists some input density operator $\sigma$ on $\hr$ with
\[
   \left\|\strut\mathcal{E}(\sigma)-|\psi\rangle\langle\psi|\,\right\|_{\rm Tr}\leq\delta.
\]
We may also assume that $N_\delta\neq\emptyset$. Let $N_\delta=:\left\{|\varphi_1\rangle,\ldots,|\varphi_N\rangle\right\}$.
Our task is to upper-bound the number of vectors $N$.
By definition, there exist density operators $\sigma_i$ on $\hr$ such that
$\|\mathcal{E}(\sigma_i)-|\varphi_i\rangle\langle\varphi_i|\,\|_{\rm Tr}\leq\delta$.
For $1\leq i\leq N$, define the projectors $P_i:=|\varphi_i\rangle\langle\varphi_i|$,
and set $P_{N+1}:=\idn-\sum_{i=1}^N |\varphi_i\rangle\langle\varphi_i|$. Let $\{|k\rangle\}_{k=1}^{\dim \hr'}$
be an orthonormal basis of $\hr'$.
Now we define a quantum operation $\mathcal{Q}$ from the trace-class operators on $\hr'$ to those on $\C^{N+1}$ via
\[
   \mathcal{Q}(a):=\sum_{i=1}^{N+1}\sum_{k=1}^{\dim\hr'} |e_i\rangle\langle k|P_i a P_i |k\rangle\langle e_i|,
\]
where $\left\{|e_i\rangle\right\}_{i=1}^{N+1}$ denotes an arbitrary orthonormal basis of $\C^{N+1}$.
It is clear that $\mathcal{Q}$ is completely positive (Kraus representation), and one easily
checks that $\mathcal{Q}$ is also trace-preserving. This is also true if $\dim\hr'=\infty$; then,
the corresponding infinite series is absolutely convergent in $\|\cdot\|_{\rm Tr}$-norm, and inherits
complete positivity from its partial sums.
Moreover, for $1\leq j\leq N$,
we have
\[
   \mathcal{Q}(P_j)=\sum_k |e_j\rangle\langle k|P_j|k\rangle\langle e_j|=|e_j\rangle\langle e_j|.
\]

Consider the equidistributed ensemble $\mathbb{E}_\sigma:=\left\{\frac 1 N,\sigma_i\right\}_{i=1}^N$, and
let $\sigma:=\frac 1 N \sum_{i=1}^N \sigma_i$. Due to the monotonicity of relative entropy with
respect to quantum operations, we have
\begin{eqnarray*}
   \chi\left(\strut\mathcal{Q}\circ \mathcal{E}(\mathbb{E}_\sigma)\right)&=&\frac 1 N \sum_{i=1}^N
   S\left(\strut \mathcal{Q}\circ\mathcal{E}(\sigma_i),\mathcal{Q}\circ\mathcal{E}(\sigma)\right)
   \leq \frac 1 N \sum_{i=1}^N S(\sigma_i,\sigma)\\
   &=& \chi(\mathbb{E}_\sigma)\leq \log d.
\end{eqnarray*}
The trace distance is also monotone with respect to quantum operations as explained in Ref.~\refcite{NielsenChuang}.
Thus, for every $1\leq i\leq N$,
\[
   \left\|\mathcal{Q}\circ\mathcal{E}(\sigma_i)-\mathcal{Q}(P_i)\right\|_{\rm Tr}
   \leq \|\mathcal{E}(\sigma_i)-P_i\|_{\rm Tr}=\|\mathcal{E}(\sigma_i)-|\varphi_i\rangle
   \langle\varphi_i|\|_{\rm Tr}\leq \delta.
\]
Let now $\Delta:=\frac 1 N \sum_{i=1}^N \mathcal{Q}(P_i)=\frac 1 N \sum_{i=1}^N |e_i\rangle\langle e_i|$,
then $S(\Delta)=\log N$, and
\[
   \left\|\mathcal{Q}\circ\mathcal{E}(\sigma)-\Delta\right\|_{\rm Tr}
   \leq \frac 1 N \sum_{i=1}^N \left\|\mathcal{Q}\circ \mathcal{E}(\sigma_i)-\mathcal{Q}(P_i)\right\|_{\rm Tr}
   \leq \delta.
\]
Now we use the Fannes inequality as given in Ref.~\refcite{NielsenChuang}. It
states\footnote{Note that the notation in Ref.~\refcite{NielsenChuang}
differs from the notation in this paper: it holds $T(\rho,\sigma)={\rm Tr}|\rho-\sigma|=\|\rho-\sigma\|_1=2\cdot\|\rho-\sigma\|_{\rm Tr}$.
} that for density matrices $\rho$ and $\sigma$ with trace distance $\|\rho-\sigma\|_{\rm Tr}\leq e^{-1}$, it holds
\[
   \left| S(\rho)-S(\sigma)\right|\leq 2\|\rho-\sigma\|_{\rm Tr}\log d +\eta(2\cdot \|\rho-\sigma\|_{\rm Tr}),
\]
where $d$ is the dimension of the Hilbert space and $\eta(x)=-x\log x\geq 0$. In our case, this inequality yields
\begin{eqnarray*}
   S\left(\strut \mathcal{Q}\circ \mathcal{E}(\sigma_i)\right)=\left| S\left(\strut
   \mathcal{Q}\circ \mathcal{E}(\sigma_i)
   \right)-S\left(\strut\mathcal{Q}(P_i)\right)\right|
   \leq 2\delta\log(N+1)+\eta(2\delta),\\
   \left| S\left(\strut \mathcal{Q}\circ \mathcal{E}(\sigma)\right)-S(\Delta)\right|
   \leq 2\delta\log(N+1)+\eta(2\delta).
\end{eqnarray*}
Altogether, we get
\begin{eqnarray*}
   \log d &\geq &\chi\left(\strut \mathcal{Q}\circ\mathcal{E}(\mathbb{E}_\sigma)\right)
   =S\left(\strut \mathcal{Q}\circ \mathcal{E}(\sigma)\right)-\frac 1 N \sum_{i=1}^N
   S\left(\strut \mathcal{Q}\circ\mathcal{E}(\sigma_i)\right)\\
   &\geq& S(\Delta)-2\delta\log(N+1)-\eta(2\delta)-\frac 1 N \sum_{i=1}^N \left(\strut
   2\delta\log(N+1)+\eta(2\delta)\right)\\
   &=&\log N - 4\delta\log(N+1)-2\eta(2\delta)\\
   &\geq& (1-4\delta)\log N - 4\delta\log 2+4\delta\log(2\delta),
\end{eqnarray*}
where we have used the inequality $\log(N+1)\leq \log N +\log 2$ for $N\geq 1$.
The claim follows by rearranging.\qed

\section{Conclusion}
We have shown in this paper that classical and quantum Kolmogorov complexity agree for classical
strings up to an additive constant.
In some sense, quantum computers are no more powerful in describing
classical strings than classical computers.

Yet, there is another, more positive way to state the result: 
This also means that quantum complexity
is an extension of classical complexity to the domain of quantum states, similar to the way that
von Neumann entropy extends classical Shannon entropy. Thus, every result on quantum complexity
will contain some classical result as a special case. Moreover, this allows to treat both classical and quantum
complexity in a single mathematical framework.

\end{document}